\documentclass[twocolumn,showpacs,preprintnumbers,amsmath,amssymb,superscriptaddress,aps]{revtex4}

\usepackage{graphicx}
\usepackage{dcolumn}
\usepackage{bm}

\begin{document}

\title{Mode hopping strongly affects observability of dynamical
      instability in optical parametric oscillators}

\author{Axelle Amon} \email{axelle.amon@univ-rennes1.fr}
\affiliation{Laboratoire PALMS, UMR CNRS 6627, Universit\'e de Rennes
I, Campus de Beaulieu, F-35042 Rennes Cedex, France}
\author{Marc Lefranc}
\affiliation{Laboratoire PhLAM, UMR CNRS 8523, CERLA, Universit\'e des
  Sciences et Technologies de Lille, F-59655 Villeneuve d'Ascq Cedex, France}

\date{\today}

\begin{abstract}
  Theoretical investigations of dynamical behavior in optical
  parametric oscillators (OPO) have generally assumed that the cavity
  detunings of the interacting fields are controllable parameters.
  However, OPOs are known to experience mode hops, where the system
  jumps to the mode of lowest cavity detuning.  We note that this
  phenomenon significantly limits the range of accessible detunings
  and thus may prevent instabilities predicted to occur above a
  minimum detuning from being evidenced experimentally. As a simple
  example among a number of instability mechanisms possibly affected
  by this limitation, we discuss the Hopf bifurcation leading to
  periodic behavior in the monomode mean-field model of a triply
  resonant OPO and show that it probably can be observed only in very
  specific setups.
\end{abstract}

\maketitle

\section{Introduction}

Continuous-wave optical parametric oscillators (OPOs) are tunable
sources of coherent light that have proved extremely useful in quantum
optics or high resolution spectroscopy~\cite{quantum97}. They also
have attracted great interest as model systems in nonlinear dynamics
because they are based on the simplest optical nonlinearity,
three-wave mixing, and are expected to exhibit complex dynamical
behavior in some regions of parameter space. Indeed, many theoretical
studies have predicted a variety of complex temporal and
spatio-temporal dynamics (see, e.g.,
\cite{neil78,drummond80,lugiato88,oppo94,staliunas95,marte98,schwob98,%
tlidi00,oppo00}).  In particular, the simplest model of a triply
resonant~OPO~(TROPO), the degenerate monomode mean-field model, was
shown twenty-five years ago to display a Hopf bifurcation leading to
periodic behavior~\cite{neil78,drummond80}, and later to exhibit
deterministic chaos ~\cite{lugiato88}. Quite surprisingly, this
instability has to our knowledge not yet been observed
experimentally. Although oscillatory behaviors have been reported in
several experiments~\cite{richy95,suret00,suret01a,suret01b}, they
have been shown to stem either from thermal
effects~\cite{suret00,suret01a} or from the interaction of transverse
modes~\cite{suret01b}. Very recently, chaotic behavior has been
evidenced in a TROPO~\cite{amon04} but is believed to be also linked
to multimode operation.

Most of the instabilities that have been investigated theoretically
require large values of the cavity detunings and of the pumping rate.
As the latter is obviously limited by the power of the pump laser, it
would only be natural to identify it as the limiting factor when
searching for the instability~\cite{oppo00}. As for detunings, they
are generally considered as parameters that can be easily tuned so as
to pull the system away from equilibrium and observe complex
behaviors. However, there are two limitations on the values cavity
detunings can take. The first is simply that the OPO threshold
increases quadratically with detuning, and thus that operation is
restricted to detuning values for which threshold power remains below
available pump power. The second arises because of a phenomenon known as
mode hopping: OPOs spontaneously choose their operation frequency so
as to operate on the cavity mode with smallest detuning. As the OPO is
pulled away from a cavity resonance in order to increase detuning, a
more favorable operating point appears near another cavity resonance
and the system jumps to it. Because mode hops in OPOs occur for
variations of cavity length as small as a few nanometers, they make it
difficult to stabilize doubly or triply resonant OPO and to achieve
smooth tuning~\cite{eckardt91,debuisschert93,henderson95}. Here, our
concern is that mode hops prevent the OPO from achieving high values
of the detunings and thus restraint the parameter range that can be
explored. Such a limitation might very well preclude the experimental
observation of dynamical behaviors predicted theoretically.

In this paper, we discuss this problem in a simple example, the Hopf
bifurcation that leads to periodic behavior in the monomode mean-field
model of a TROPO. We show that mode selection in OPOs indeed prevents
experimental observation of this instability in most practicable
configurations. We first recall the basic properties of the monomode
mean-field TROPO model and the conditions under which the Hopf
bifurcation occurs. In the limit of infinite pump power, a simple
lower bound for signal detunings at which the Hopf bifurcation can
occur is easily obtained, which confirms that this bifurcation
requires high values of the signal detuning. In a second part we
derive the expression of the maximal value that signal detuning can
reach before a mode hop occurs, which depends on the length and
finesse of the cavity as well as on crystal birefringence. By
comparing the two bounds and searching for parameter regions where
they are compatible, we find that mode hops generally keep the TROPO
away from parameter ranges where the Hopf bifurcation can be observed,
unless very high-finesse and very short cavities are used, which would
make operation extremely difficult. This is confirmed by a numerical
exploration at finite pump power of the parameter space of this model
for various values of the cavity finesses. It shows that mode hopping
rather than pump power is the limiting factor in order to reach
instability. This result provides a plausible explanation of the fact
that the Hopf bifurcation of the monomode TROPO has not yet been
observed experimentally. It also calls for further investigations in
order to determine whether mode hopping also interferes with other
predicted instabilities.

\section{TROPO monomode mean-field degenerate model: Hopf instability}
\label{sec:hopf}

We now recall the main features of the simplest TROPO model, the
degenerate longitudinally and transversely monomode mean-field
model~\cite{neil78,drummond80,lugiato88}. Light generation in an
optical parametric oscillator is based on parametric down-conversion
in a nonlinear crystal of a pump photon into two lower-frequency
photons called signal and idler. In a TROPO, the optical cavity
enclosing the crystal is resonant for all three fields so as to
minimize operation threshold. In the mean-field (a.k.a. uniform-field)
approximation, the time evolution of the normalized amplitudes $A_s$,
$A_i$ and $A_p$ of the signal, idler and pump fields inside the cavity
can be described by the following differential
equations~\cite{lugiato88}:
\begin{subequations}
\begin{eqnarray}
  \dot{A}_s &=& -(1+i\Delta_s)A_s + A_i^*A_p,\label{modelthreesig}\\
  \dot{A}_i &=& -(1+i\Delta_i)A_i + A_s^*A_p,\label{modelthreeid}\\
  \dot{A}_p &=& \gamma
\left[-(1+i\Delta_p)A_p - A_sA_i  + E\right],
\end{eqnarray}
\label{modelthree}
\end{subequations}
where $\Delta_s$, $\Delta_i$ and $\Delta_p$ are the detunings between
the optical frequency and the frequency of the closest cavity
resonance for the signal, idler and pump fields, respectively, and $E$
is the pumping rate. The time unit is the cavity decay time of the
signal field and $\gamma$ is the cavity decay rate for the pump. In
this paper we focus on the stationary regimes of
Eqs~\eqref{modelthree}, and are interested in determining when they
become unstable to give birth to periodic oscillations depending on
values of control parameters.

A little known property of Eqs.~\eqref{modelthree} is that although
the signal and idler fields are in principle distinct, their time
evolution can be described by a single amplitude after transients have
died out. First it should be noted that stationary solutions of
Eqs.~\eqref{modelthree} exist only for $\Delta_s = \Delta_i$, a
relation that can be shown to follow from photon number
conservation~\cite{Fabre97}. Replicating a similar calculation carried
out in the analysis of a bimode model~\cite{amon03}, it is then easy
to show that Eqs.~\eqref{modelthreesig} and \eqref{modelthreeid} imply
that after a sufficiently long time, the amplitudes $A_s$ and $A_i$
are equal up to a constant phase difference that can always be removed
by a redefinition of the amplitudes. This is obviously linked to the
fact that signal and idler photons are twin photons created in the
same quantum process. Without loss of generality, the asymptotic
dynamics of the TROPO can then be modeled by the following normalized
equations (degenerate model) describing the time evolution of the
complex amplitude of the signal field $A_s$ and of the pump field
$A_p$~\cite{lugiato88}:
\begin{subequations}
\begin{eqnarray}
  \dot{A}_s &=& -(1+i\Delta_s)A_s + A_s^*A_p,\\
  \dot{A}_p &=& \gamma
\left[-(1+i\Delta_p)A_p - A_s^2  + E\right],
\end{eqnarray}
\label{model}
\end{subequations}
 For pumping rates above the parametric
emission threshold given by
\begin{eqnarray}
E_{th}^2 = (1+\Delta_p^2)(1+\Delta_s^2),
\label{seuil}
\end{eqnarray}
equations~\eqref{model} have non-zero stationary solutions which have
been shown to fit accurately experimental observations in the vicinity
of threshold~\cite{richy95}. When pump rate is increased, these
stationary solutions can become unstable through a Hopf bifurcation
giving rise to oscillatory
behavior~\cite{neil78,drummond80,lugiato88}. A necessary condition for
this bifurcation is~\cite{lugiato88}
\begin{equation}
  \label{cond_hopf}
\Delta_p \Delta_s < - \left[ 1+\frac{\gamma (1 +
\Delta_p^2)}{2}\right],
\end{equation}
which ensures that there is a finite pump rate $E_H > E_{th}$ at which
the stationary nonzero solution bifurcates to a periodic
solution, which is is given by~\cite{lugiato88}
\begin{eqnarray}
E_H^2 &=& \left[ \frac{\gamma^2 (1+\Delta_p^2) +
  4(1+\gamma)}{-2(1+\gamma)^2[1+\frac{2(1+\Delta_p \Delta_s)}{\gamma
  (1 + \Delta_p^2)}]} -(\Delta_p \Delta_s - 1)\right]^2 \nonumber \\ &
  & + (\Delta_p+\Delta_s)^2,
\label{hopf}
\end{eqnarray}
At higher pump rates, the limit cycle born in the Hopf bifurcation
undergoes a period-doubling cascade leading to chaos~\cite{lugiato88}.

As inequality~\eqref{cond_hopf} is closer and closer to equality, the
Hopf threshold $E_H$ given by~\eqref{hopf} becomes larger and larger
and is eventually rejected to infinity. For a given maximal pump rate
available, $E_{\max}$, whether the instability can be observed at fixed
detunings inside the parameter domain delimited by~\eqref{cond_hopf}
is determined by the inequality $E_{\max} >
E_H(\Delta_p,\Delta_s)$. Since we are interested here in specifying
the unstable region by simple bounds on the detunings, we first assume
that we have infinite pump power available. Under this approximation,
whose validity will be checked in numerical simulations in
Sec.~\ref{finite}, the Hopf instability domain is solely determined by
inequality~\eqref{cond_hopf}.

\begin{figure}
\includegraphics[scale=0.25]{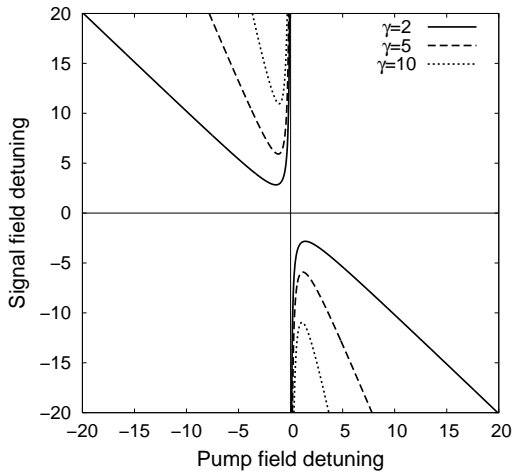}
\caption{\label{courbes} Map ($\Delta_p, \Delta_s$) for different
  values of the parameter $\gamma$. The curves delimits the areas for
  which condition~\eqref{cond_hopf} is fulfilled.}
\end{figure}

The instability domains in the $(\Delta_p,\Delta_s)$ plane have been
plotted in Fig.~\ref{courbes} for different values of $\gamma$, and
are seen to be bounded away from the origin. Along their boundaries,
\eqref{cond_hopf} is an equality and defines the signal detuning as a
function of pump detuning. It is easily found that the minimum
absolute value that the signal detuning can take on the boundary, and
hence in the instability domain, is:
\begin{equation}
  \label{eq:deltamin}
\Delta_{\min}^H = \min\{|\Delta_s|\} = \sqrt{\gamma (\gamma + 2)},
\end{equation}
and is obtained for $|\Delta_p| = \sqrt{(2 + \gamma)/\gamma}$. Note
that $\Delta_{\min}^H$ is roughly proportional to $\gamma$ and thus
increases with it.

Since~\eqref{cond_hopf} still holds at finite pump power, the lower
bound provided by~\eqref{eq:deltamin} is always valid. This clearly
shows that the Hopf bifurcation of the monomode degenerate mean-field
model can only occur for sufficiently high signal detunings. In the
next Section, we describe the process of mode hopping and how it
limits the values that signal detunings can take leading to a maximal
absolute value for the detunings. Whether the two constraints can be
satisfied simultaneously will eventually determine whether the Hopf
bifurcation can be observed experimentally.

\section{Limitation of signal detuning due to mode hopping}
\label{hopping}

\subsection{Theory}
\label{sec:theory}

In the theoretical analysis of Sec~\ref{sec:hopf}, it was assumed that
the frequency detuning of the signal field is a fixed parameter. This
is however not entirely true, because the operating frequency of an
OPO is not actually chosen by the experimentalist but results from a
complex mode selection process. While small variations of the cavity
length most often modify frequency and hence detuning gradually,
sudden jumps will occur when a remote operating point becomes more
favorable.  As we discuss below, this process tends to minimize
frequency detuning and consequently to limit its maximum value, which
can be derived analytically.

As with any optical oscillator where an amplifying medium is enclosed
inside an optical cavity, there are two main constraints that
determine the operating conditions of an OPO: the first one is that
the gain in the amplifying medium must be sufficient to overcome
cavity losses over each round trip in the cavity, the second one is
that the generated field must be nearly resonant with one of the
cavity modes so that amplification by the gain medium is cumulative
over successive round trips.

In optical parametric oscillators, the nonlinear interaction is
optimal when the relative phases of the three interacting waves remain
fixed during propagation. The phase-matching condition is easily
formulated in a corpuscular point of view: the conversion of one pump
photon into signal and idler photons must satisfy energy and momentum
conservation:
\begin{subequations}
    \label{eq:conservation}
    \begin{eqnarray}
    \omega_p &=& \omega_s+\omega_i\label{eq:energy_conservation}\\
    \vec{k}_p &=& \vec{k}_s+\vec{k}_i\label{eq:momentum_conservation}
  \end{eqnarray}
\end{subequations}
where $\omega_{p,s,i}$ and $k_{p,s,i}$ denote the frequencies and the
wavevectors of the pump, signal and idler fields.

The pump properties being fixed, relations~\eqref{eq:conservation}
generally single out unique values for the signal and idler
frequencies and much of OPO design consists in ensuring that these
values fall in the desired frequency range. However, while energy
conservation~\eqref{eq:energy_conservation} strictly holds for
continuous-wave OPOs, momentum conservation may be satisfied only
approximately because of the finite size of the nonlinear crystal.
Thus there is a small but finite frequency domain around the exact
phase-matching frequency where the nonlinear gain is adequate for OPO
operation. It is usually the case that many cavity resonances fall
inside this domain, and an important problem is to determine the
resonances near which oscillation will actually occur.

Monomode optical parametric oscillators behave as homogenously
broadened lasers in that when there are several frequencies for which
gain overcome losses in the empty cavity, the one with the lowest
threshold takes over by saturating the gain in such a way that
competing modes remain below threshold. Mode selection thus amounts to
determining which operating modes have lowest threshold. If variation
of raw gain in the domain around perfect phase-matching is neglected,
and considering pump detuning as a fixed parameter, this is equivalent
to finding the allowed operating frequencies for signal and idler
fields that have lowest frequency detuning, as
expression~\eqref{seuil} for parametric threshold shows.

In doubly and triply resonant OPOs, this problem is made difficult by
the fact that the signal and idler fields must be simultaneously
resonant. Thus operation can only take place at coincidences between
two frequency combs. In general the two combs have different periods,
either because the signal and idler fields have very different
frequencies or because they are polarized along different axes of a
birefringent crystal (the so-called type-II phase-matching~\cite{boyd}). A
singular configuration that we will not consider here is when signal
and idler have identical polarizations and frequencies (type-I phase
matching at degeneracy). As a result, frequency tuning in OPOs is a
complicated problem that has been studied very carefully both in the
type-I and type-II cases~\cite{eckardt91,debuisschert93}.

The detunings of the signal and the idler fields are given respectively
by $\Delta \omega_s = \omega_s - \omega_s^R$ and $\Delta \omega_i =
\omega_i - \omega_i^R$, where $\omega_s$ and $\omega_i$ are the
optical frequencies of the signal and idler fields and $\omega_s^R$
and $\omega_i^R$ are the closest resonance frequencies of the cavity.
For the sake of simplicity, we assume in the following that $\omega_s$
and $\omega_i$ are close to a 1:1 ratio but the argument can be
generalized easily to any rational number.

Taking energy conservation \eqref{eq:energy_conservation} into
account, the total frequency mismatch $\Delta \omega = \Delta \omega_s
+ \Delta \omega_i$ for the signal-idler mode pair is given by :
\begin{equation}
\label{destot}\\
\Delta \omega = \omega_p - \omega_s^R - \omega_i^R,
\end{equation}
which does not depend of the specific oscillation frequencies
$\omega_s$ and $\omega_i$ chosen by the system, but only of the pump
frequency and the resonance frequencies of the cavity for the signal
and the idler modes. For our purposes, $\Delta \omega$ is the relevant
quantity to consider as the individual detunings $\Delta \omega_s$ and
$\Delta \omega_i$ are proportional to it in a stationary state, as a
consequence of energy and photon number conservation~\cite{Fabre97}. The mode
pair selected by the OPO will be the one that minimizes the total
frequency mismatch $\Delta \omega$ so as to minimize threshold.

Taking into account that $\omega_s^R$ and $\omega_i^R$ belong to two
frequency combs specified by the free spectral ranges $\delta
\omega_s$ and $\delta \omega_i$:
\begin{equation}
  \label{eq:modenumber}
  \omega_s^R = N_s \delta\omega_s,\quad \omega_i^R = N_i \delta\omega_i
\end{equation}
where $N_s$ and $N_i$ are the mode indices, and having in mind that
$|\delta \omega_s - \delta \omega_i| \ll \delta \omega_s,\delta
\omega_i$, we rewrite~\eqref{destot} as

\begin{equation}
  \label{eq:clusteradj}
  \Delta \omega = \omega_p -
     \bar{N}(\delta\omega_s+\delta\omega_i)-
    \Delta N (\delta\omega_s-\delta\omega_i)
\end{equation}
with $\bar{N}=(N_s+N_i)/2$ and $\Delta N = (N_s-N_i)/2$. Under our
hypotheses, we have $\Delta N \ll \bar{N}$. Hence the last term
in~\eqref{eq:clusteradj} is in first approximation negligible compared
to the second term, and the integer value of $\bar{N}$ that minimizes
$\Delta\omega$ is determined independently of $\Delta N$. Then the
optimization problem can be refined by searching for the integer value
of $\Delta N$ that minimizes~\eqref{eq:clusteradj} at fixed $\bar{N}$.
Studies of tuning properties of double or triply resonant OPO have
shown that the operating modes are grouped into clusters, each cluster
consisting of a sequence of adjacent modes
\cite{eckardt91,debuisschert93}. In~\eqref{eq:clusteradj}, $\bar{N}$
indicates the cluster and $\Delta N$ distinguishes between adjacent
modes inside the cluster.

An important consequence of~\eqref{eq:clusteradj} is that $\Delta
\omega$ can at best be adjusted in steps of
$|\delta\omega_s-\delta\omega_i|$. With such steps, the frequency
mismatch $\Delta \omega$ can always be made to belong to the interval
$[-\frac{|\delta \omega_s - \delta \omega_i|}{2},\frac{|\delta
  \omega_s - \delta \omega_i|}{2}]$ but cannot be brought closer to
zero. This minimal $\Delta \omega$ corresponds to the mode pair chosen
by the OPO. By considering the worst case, the maximal frequency
mismatch that can be reached is
\begin{equation}
\Delta \omega_{\max} = \frac{|\delta \omega_s - \delta \omega_i|}{2},
\label{dwmax}
\end{equation}
which agrees with the general expression of the detuning given
in~\cite{eckardt91}. As the OPO is pulled away from an exact cavity
resonance by increasing the cavity length, the frequency mismatch will
increase up to the maximal value given by~\eqref{dwmax}, at which
point there will be an operating mode with a lower mismatch to which
the OPO will switch. This phenomenon is well known as mode
hopping~\cite{eckardt91,debuisschert93}.

Before~\eqref{dwmax} can be compared with the bound found for the Hopf
bifurcation in Sec.~\ref{sec:hopf}, it has to be expressed in the same
units. The detunings used in equations~\eqref{model} are normalized so
that half-height width of the cavity resonance is 2. Since the cavity
finesse $\mathcal{F}$ is defined so that the half-height width is
$\delta \omega/\mathcal{F}$, we have
\begin{subequations}
  \label{eq:normalizeddetunings}
\begin{eqnarray}
\Delta_s &=& 2 \mathcal{F}_s \frac{\Delta \omega_s}{\delta
\omega_s},\\ \Delta_i &=& 2 \mathcal{F}_i \frac{\Delta
\omega_i}{\delta \omega_i}.
\end{eqnarray}
\label{norme}
\end{subequations}

As mentioned before, $\Delta_s$ and $\Delta_i$ are not independent in
the stationary regime but obey the simple relation $\Delta_s =
\Delta_i$ as a consequence of energy and photon number
conservation~\cite{Fabre97}. Taking into account that $\Delta\omega =
\Delta\omega_s+\Delta\omega_i$, one obtains
from~\eqref{eq:normalizeddetunings}:
\begin{equation}
\Delta_s = \Delta_i = \frac{2\mathcal{F}_s \mathcal{F}_i \Delta
  \omega}{\mathcal{F}_i \delta \omega_s +
  \mathcal{F}_s \delta \omega_i}.
\end{equation}
This expression was derived assuming $\omega_s\simeq\omega_i$, so that
we should fix $\mathcal{F}_s \simeq \mathcal{F}_i$ for consistency.
The maximal value of the frequency mismatch authorized by mode
selection, as given by equation~\eqref{dwmax}, can thus be rewritten
as
\begin{eqnarray}
\Delta_{\max}^M = \mathcal{F}_s \frac{|\delta \omega_s - \delta
 \omega_i|}{\delta \omega_s + \delta \omega_i},
\label{eckardt}
\end{eqnarray}

Before we can discuss whether the Hopf bifurcation can be observed in
typical experiments, we have to reformulate~\eqref{eckardt} in terms
of the experimental configuration.

Given that
\begin{equation}
  \delta \omega_{s,i} = \frac{2 \pi c}{2(L_{cav} +
  (n_{s,i} - 1) l_c)}\label{eq:fsr}
\end{equation}
where $L_{cav}$ is the geometric length of the cavity, $l_c$ is the
length of the nonlinear crystal, $n_{s,i}$ the indices of the signal
(resp. idler) fields and $c$ is the celerity of light in the vacuum,
\eqref{eckardt} can be rewritten as
\begin{eqnarray}
\Delta_{\max}^M = 2\mathcal{F}_s \frac{|\delta n| l_c}{[L]},
\label{debui}
\end{eqnarray}
where
\begin{displaymath}
[L] = 2(L_{cav} + (\frac{n_s + n_i}{2} - 1) l_c)
\end{displaymath}
is the average optical path for the signal and the idler fields over one
round trip in the cavity and
\begin{displaymath}
  \delta n = \frac{|n_s - n_i|}{2}.
\end{displaymath}

In the limit case of a monolithic OPO ($L_{cav}=l_c$),
expression~\eqref{debui} leads to the remarkably simple expression
\begin{equation}
  \label{eq:monolithic}
  \Delta_{\max}^M/\mathcal{F}_s = \frac{2|\delta n|}{n_s+n_i}
\end{equation}
showing the key role played by cristal birefringence.
Expression~\eqref{eq:monolithic} yields an absolute upper bound
for~\eqref{debui} since cavity length obviously cannot be smaller than
crystal length.

Note that since $\delta n \ll n_{s,i}$ for a standard crystal,
expressions~\eqref{debui} and~\eqref{eq:monolithic} ensure that the
small-detuning hypothesis of the mean-field approximation is
fulfilled for the signal and idler fields.

We are now in a position to obtain a simple criterion to determine
whether the Hopf bifurcation can be observed in a given configuration
at infinite pump power.
Obviously, we must have
\begin{equation}
  \label{eq:criteriontrivial}
  \Delta_{\min}^H < \Delta_{\max}^M
\end{equation}
which, using~\eqref{eq:deltamin} and~\eqref{debui}, and expressing the
pump cavity decay rate as
\begin{displaymath}
  \gamma = \frac{\mathcal{F}_s}{\mathcal{F}_p},
\end{displaymath}
translates into
\begin{equation}
  \label{eq:amonlefranc}
  \sqrt{
    \frac{1}{\mathcal{F}_p}
    \left(
      \frac{1}{\mathcal{F}_p}+
      \frac{2}{\mathcal{F}_s}
    \right)}
  <
2 \frac{|\delta n| l_c}{[L]}  
\end{equation}
which is the main result of our paper. Again,
\eqref{eq:amonlefranc} simplifies in the monolithic case. A noteworthy
feature of inequality~\eqref{eq:amonlefranc} is the asymmetry in the
dependences with respect to the pump and signal finesses. It is easily
seen that increasing pump finesse is much more effective to
have~\eqref{eq:amonlefranc} satisfied. Assuming $\mathcal{F}_p=50$ and
$\mathcal{F}_s=500$, doubling the pump finesse decreases the left hand
side of~\eqref{eq:amonlefranc} by $46\%$ while doubling the signal
finesse only does so by $4\%$.

\subsection{Numerical estimates}
\label{sec:numerical-estimates}

To get a better understanding of whether the criterion obtained in
Sec~\ref{sec:theory} is easily satisfied or not, we now compute
numerical estimates for typical experimental configurations. In
previous experiments~\cite{suret00,suret01a,suret01b,amon03,amon04},
we used a KTP crystal of length $l_c = 15$ mm cut for type-II
phase-matching, with an extraordinary index $n_e = 1.75$ and an
ordinary index $n_o = 1.83$. The crystal is enclosed in a Fabry-Perot
cavity delimited by two spherical mirrors with a radius of curvature
of 5 cm. The cavity finesses for the signal and pump fields are around
500 and 50, respectively.

\begin{figure}
\includegraphics[scale=0.30]{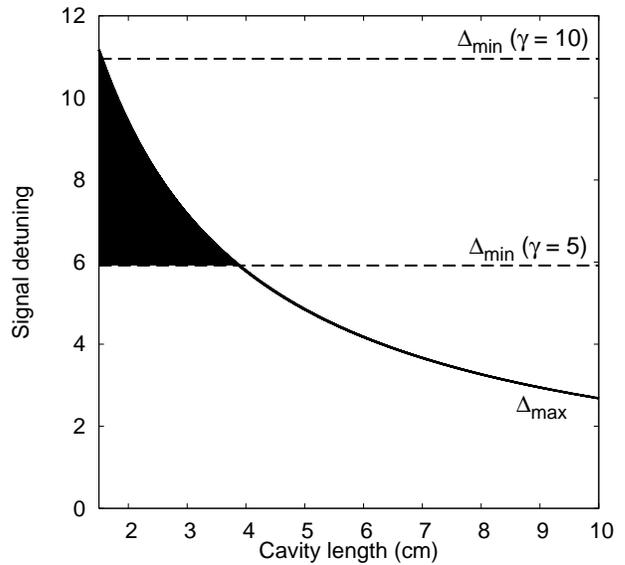}
\caption{\label{max} Solid lines : values of maximal detuning
  authorized by mode selection when cavity length is varied
  between 1.5 cm and 10 cm. Dashed lines : minimal detuning permitted
  by the conditions of the Hopf bifurcation for two different values
  of $\gamma$ : 
  5 and 10.}
\end{figure}

The most easily adjustable parameter is the geometrical cavity length,
which is bounded from below by the crystal length (1.5 cm) and from
above by the concentric condition ($\sim$10 cm). Figure~\ref{max}
shows the evolution of the maximal detuning $\Delta_{\max}^M$ with
cavity length. The main feature is that $\Delta_{\max}^M$ decreases
monotonically with cavity length, as is easily seen in
expression~\eqref{debui}. Thus the most favorable situation will
always be obtained in the ``monolithic'' configuration where cavity
length equals crystal length. In practice, this configuration cannot
be reached when spherical mirrors are used because of the size of the
crystal mount, but the shortest feasible cavity length (3 cm in our
experiments) should be sought.

In our experiment, $\gamma = \mathcal{F}_s/\mathcal{F}_p=10$, from
which we can compute the value $\Delta_{\min}^H = \sqrt{\gamma (\gamma
  + 2)}$ of the minimal detuning necessary to obtain the Hopf
bifurcation. This value does not depend on cavity length and is
represented in Fig.~\ref{max} by the upper dashed horizontal line. We
see that there is hardly a configuration where
$\Delta_{\min}^H<\Delta_{\max}^M$. In this configuration, the Hopf
bifurcation cannot be observed regardless of pump power because mode
hopping prevents to reach sufficiently high detunings.

In order to lower $\Delta_{\min}^H$, assume now that pump finesse is
increased to 100 so that $\gamma$ decreases to 5. This corresponds to
the lower horizontal dashed line in Fig.~\ref{max}. There is now a
range of cavity lengths where $\Delta_{\min}^H<\Delta_{\max}^M$ as
shown by the black area between the two curves.

If $\mathcal{F}_s$ and $\mathcal{F}_p$ are increased simultaneously
keeping $\gamma$ constant, $\Delta_{\max}^M$ will increase while
$\Delta_{\min}^H$ remains unchanged, and the instability region will
also widen.

\subsection{Conclusion}

In this section we have taken into account the fact that the signal
frequency detuning is not a fixed parameter but is determined by the
operating frequency chosen by the OPO so as to minimize this detuning.
As a result the signal frequency detuning is bounded from above by the
mode hopping phenomenon. Since signal detunings at which periodic
behaviors appear are bounded from below, the Hopf bifurcation can only
be observed if $\Delta_{\min}^H < \Delta_{\max}^M$. Our analysis has
shown that even in the ideal case where infinite pump power is
available, mode hopping can prevent from reaching detunings
sufficiently large to observe the Hopf bifurcation of the monomode
model.

The numerical estimates obtained for the configuration used in our
previous experiments explain why we could not observe the Hopf
bifurcation in this setup. In order to obtain instabilities, one
should take into account that $\Delta_{\min}^H$ depends on the ratio
of signal and pump finesses while $\Delta_{\max}^M$ is proportional to
signal finesse. More generally, the finesses should satisfy
inequality~\eqref{eq:amonlefranc}. As will be discussed in
Sec.~\ref{setup}, the theoretically most favorable configurations are
extremely difficult to build experimentally. In all cases, cavity
should be made as short as possible.

It remains to be checked that in cases where the instability can be
observed, it persists when pump power is limited. We do so in Section
\ref{finite}.

\section{Numerical investigations at finite pump power}
\label{finite}

So far, our analysis has assumed infinite pump power, which has
allowed us to obtain simple analytical formulas such
as~\eqref{eq:amonlefranc} to decide whether the Hopf bifurcation is
screened by mode hopping or not. We now have to determine how good
this approximation is in the real-life situation where only finite
pump power is available and if conclusions drawn from our theoretical
analysis remain relevant.

At finite pump power, whether the Hopf bifurcation can be observed no
longer depends only on inequality~\eqref{cond_hopf} but also on
whether the Hopf threshold~\eqref{hopf} can be reached given available
pump power. Because expression~\eqref{hopf} is much more complicated
than inequality~\eqref{cond_hopf}, we restrict ourselves here to a
numerical exploration of the detuning ranges where periodic behavior
is found and compare our results with predictions from the infinite
pump power analysis. This exploration is carried out for several
values of the cavity finesses, other parameters being chosen so as to
match our previous
experiments~\cite{suret00,suret01a,suret01b,amon03,amon04}. As we
shall see, it will allow us to conclude that pump power is not a
limiting factor with commonly available pump lasers and that the
criteria derived in the infinite pump power analysis remain relevant.

A few general observations are in order before we present our
numerical results. The relevant criterion is whether the minimum
detuning $\Delta_{\min}^H$ at which Hopf bifurcation occurs is smaller
than the larger detuning $\Delta_{\max}^M$ allowed by mode hopping.
Thus it is interesting to comment on how these bounds evolve when pump
power is limited.

Regarding onset of periodic behavior, it should be recalled that
inequality~\eqref{cond_hopf} holds regardless of pump power, that
equality can be achieved only for infinite pump power and that
otherwise the Hopf threshold~\eqref{hopf} yields a more stringent
condition than~\eqref{cond_hopf} on the detunings. As a result, the
instability regions are systematically shifted towards higher values
of the detunings, leading to an increase of $\Delta_{\min}^H$ compared
to expression~\eqref{eq:deltamin}.

As for the maximal value $\Delta_{\max}^M$ of the detuning allowed by
mode hopping [Expr.~\eqref{eckardt}], it does not depend on pump power
as it is obtained by considering the frequency combs of the cavity
resonances. However a limitation that has to be taken into account at
finite pump power is that if the OPO is below parametric emission
threshold at $\Delta_s=\Delta_{\max}^M$, then the latter bound
certainly cannot be achieved and the actual bound will be lower (the
stationary ON state must exist for the bifurcation to occur). Using
the expression for parametric threshold~\eqref{seuil}, the expression
for the maximal signal detuning value becomes
\begin{equation}
  \label{eq:newmaximal}
  \Delta_{\max}^{MT} = \min\left(\Delta_{\max}^M,
  \sqrt{\frac{E^2}{1 + \Delta_p^2}- 1}\right)
\end{equation}
where $E$ is the pump parameter and $\Delta_p$ is the pump detuning.
The compatibility between the values of the minimal detuning at which
periodic behavior occurs and of the maximal detuning at which
stationary parametric emission occurs is now more difficult to analyze
because both depend on pump detuning. However, it is easy to see
that as pump power is decreased, the former can only increase and the
latter only decrease so that the Hopf bifurcation is necessarily
harder to observe in the finite pump power case. We now assess by how
much by carrying out numerical simulations.

\begin{figure}
\includegraphics[scale=0.25]{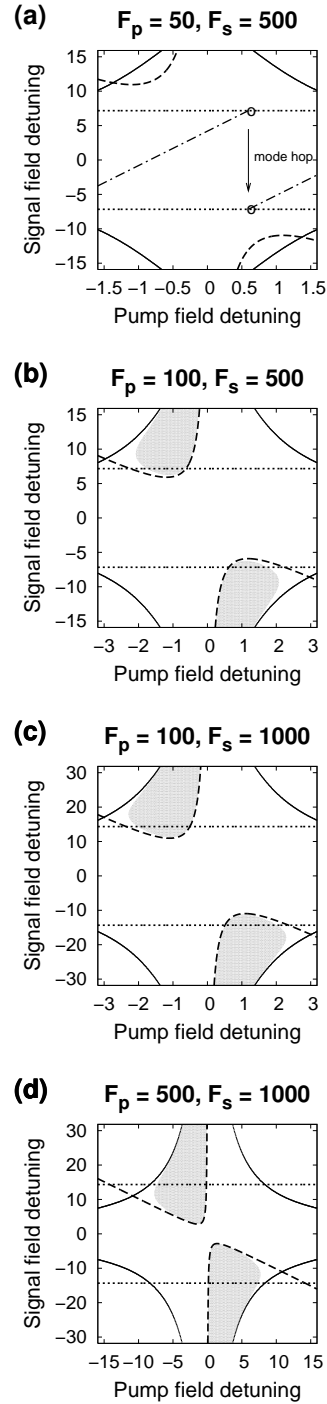}
\caption{\label{maps} Phase diagrams in the ($\Delta_p, \Delta_s$)
  plane showing the instability regions as black areas for different
  values of the finesses of the pump ($\mathcal{F}_p$) and of the
  signal ($\mathcal{F}_s$) field: (a) $\mathcal{F}_p = 50,
  \mathcal{F}_s = 500$, (b) $\mathcal{F}_p = 100, \mathcal{F}_s =
  500$, (c) $\mathcal{F}_p = 100, \mathcal{F}_s = 1000$, (d)
  $\mathcal{F}_p = 500, \mathcal{F}_s = 1000$. Solid curves
  delimit areas above parametric emission threshold. Dashed lines
  indicate where~\eqref{cond_hopf} is an equality and delimit unstable
  areas in the infinite pump power approximation.  Dotted horizontal
  lines indicate the maximal detuning value allowed by mode hops. In
  (a), the dotted and dashed line shows a possible path followed by
  detunings as cavity length is scanned. The discontinuity occurring
  as the path reaches the horizontal dotted line corresponds to a mode
  hop.}
\end{figure}

The phase diagrams in Fig.~\ref{maps} shows in black the regions in
the $(\Delta_p,\Delta_s)$ parameter plane where periodic behavior is
found for different values of the cavity finesses, the values of the
remaining parameters being fixed so as to match our
experiments~\cite{suret00,suret01a,suret01b,amon03,amon04}. For each
parameter set, the dynamical regime is classified as periodic when 
\begin{itemize}
\item Inequality~\eqref{cond_hopf} is satisfied, and 
\item $E_H(\Delta_p,\Delta_s)<E_{max}(\mathcal{F}_p,\mathcal{F}_s)$
\end{itemize}
where $E_{\max}(\mathcal{F}_p,\mathcal{F}_s)$ is the maximum pump
parameter corresponding to the pump power available in our
experimental conditions (4
W)~\footnote{$E_{max}(\mathcal{F}_p,\mathcal{F}_s) =
\sqrt{\mathcal{P}_{max}/\mathcal{P}_{th}(\mathcal{F}_{p},\mathcal{F}_{s})}$
with $\mathcal{P}_{max}$ the maximum pump power available and
$\mathcal{P}_{th}(\mathcal{F}_{p},\mathcal{F}_{s})$ the pump power at
threshold. It is known that $\gamma_p
\mathcal{P}_{th}(\mathcal{F}_{p},\mathcal{F}_{s}) \propto \gamma_s^2
\gamma_p^2$~\cite{yariv66,debuisschert93} and consequently
$\frac{E^2_{max}(\mathcal{F}_{p},\mathcal{F}_{s})}{\mathcal{F}_{p}
\mathcal{F}_{s}^2} = cst$. In our experimental setup,
$\mathcal{F}_{p0} = 45$, $\mathcal{F}_{s0} = 550$, $\mathcal{P}_{max}
\simeq 4$ W and $\mathcal{P}_{th}(\mathcal{F}_{p0},\mathcal{F}_{s0})
\simeq 10$ mW, which leads to
$\frac{E^2_{max}(\mathcal{F}_{p},\mathcal{F}_{s})}{\mathcal{F}_{p}
\mathcal{F}_{s}^2} \simeq 400.$}. Since the unstable zones are
enclosed inside the regions where parametric emission occur and the
instability regions at infinite pump power, the boundaries of these
regions, which can be computed analytically, are also shown in
Fig.~\ref{maps} so that we can estimate how well they approximate the
numerical results. In order to make meaningful comparisons between
setups corresponding to different values of the finesses, the domains
of variations of the pump and signal detunings are chosen so that
$\Delta_{s,p}\in[-\mathcal{F}_{s,p}/(10\pi),
\mathcal{F}_{s,p}/(10\pi)]$. This corresponds in each case to the same
variation of the physical cavity length ($\Delta L=\lambda/40 \pi$)
and thus there is no difference in the scans from an experimental
point of view. Since $|\Delta_{s,p}| \ll 2 \mathcal{F}_{s,p}$, this
also ensures that the small-detuning hypothesis of the mean-field
model is satisfied.

Fig.~\ref{maps}(a) shows that for parameter values corresponding to
our experimental setup, periodic behavior does not occur, explaining
why we do not observe the Hopf bifurcation in this setup. It also
shows that the boundaries of the unstable region at infinite pump
power are located outside the central band where
$\Delta_s \in [-\Delta_{\max}^M,\Delta_{\max}^M]$ and thus that mode
hopping prevents the bifurcation from being observed in this
configuration even if infinite pump power was available. The proximity
of the two curves corresponding to parametric threshold and
bifurcation at infinite pump power probably explains why no
instability can be observed.

In contrast with Fig.~\ref{maps}(a), Figs.~\ref{maps}(b) to
\ref{maps}(d) display larger and larger instability zones as cavity
finesses are increased. While in Fig.~\ref{maps}(b) the intersection
of the unstable region with the central band of allowed signal
detunings is very small, it becomes sizable in Fig.~\ref{maps}(d).
More precisely, the fraction of the central band occupied by unstable
regions is $1.6\%$ in Fig.~\ref{maps}(b), $3.5\%$ in
Fig.~\ref{maps}(c) and $17 \%$ in Fig.~\ref{maps}(d). These numbers
are meaningful as probability estimates if we assume that there is no
correlation between the signal and pump detunings on average, i.e.,
that the entire allowed band may be explored over several experiments.
It should be noted that during a scan of cavity length through a
single resonance of the pump, pump and signal detuning will vary
jointly as illustrated in Fig.~\ref{maps}(a) and that this may affect
the probability of hitting the unstable zone. However, it is expected
that there is no relation between paths followed in the
$(\Delta_p,\Delta_s)$ plane for different pump resonances so that no
part of the allowed band should remain inaccessible.

Examination of Figs.~\ref{maps}(a)-(d) shows that although increasing
finesses globally makes it easier to observe the Hopf bifurcation,
finesses for the pump and signal fields have different influences. The
twofold increase in $\mathcal{F}_p$ between Figs.~\ref{maps}(a) and
(b) clearly modifies the phase diagram much more than the twofold
increase in $\mathcal{F}_s$ between Figs.~\ref{maps}(b) and (c).
Similarly, increasing $\mathcal{F}_p$ from 100 to 500 is critical to
have a significant probability of observing the Hopf bifurcation. This
is consistent with the discussion of criterion~\eqref{eq:amonlefranc}
in Sec.~\ref{hopping} which showed that increasing pump finesse was
much more effective than increasing signal finesse. However increasing
pump finesse is extremely difficult from a practical point of view, as
we discuss in Sec.~\ref{setup}. Interestingly, it was similarly noted
in Ref.~\cite{oppo00} that increasing pump finesse while keeping
signal finesse allowed one to observe an hexagonal transverse pattern
at lower signal detunings.

An important conclusion that can be drawn from the numerical
exploration summarized in Figs.~\ref{maps}(a)-(d) is that except in
Fig.~\ref{maps}(a), the theoretical analysis at infinite pump power
provides a very good approximation of the finite pump power case, all
the better as finesses are higher and as observing the Hopf
bifurcation becomes more plausible. Indeed the instability zones are
tightly delimited by the boundaries obtained at infinite pump power
and by the parametric threshold line except for larger pump detunings.
In particular, the agreement is excellent at the tip of the unstable
zones, near the point of minimal signal detuning. This makes us
confident that for pump lasers currently available, the criterion
obtained in~\eqref{eq:amonlefranc} is effective in assessing the
probability of occurrence of the Hopf bifurcation. It also shows that
pump power is not a limiting factor as when there is a nonempty
intersection between the unstable zone and the central band of allowed
detunings, its area only marginally increases from the finite to the
infinite pump case, mostly in regions far from the minimum detuning
value.

To conclude, numerical calculations at finite pump power show that
while the Hopf bifurcation becomes harder to observe than in the
infinite pump power case, the mode hopping phenomenon remains the main
limiting factor by limiting the range of values signal detuning can
take. In order to obtain experimental evidence of the bifurcation,
building a dedicated setup would be much more effective than
increasing pump power. However, there are some experimental
difficulties in doing so, which we discuss in Section~\ref{setup}.

\section{Experimental considerations}
\label{setup}

The theoretical analysis at infinite pump power of Sec.~\ref{hopping}
and the numerical computations at finite pump power of
Sec.~\ref{finite} have suggested that the Hopf bifurcations of the
monomode mean-field model might be observable in some configurations.
In this Section, we discuss the feasibility of an experimental setup
specially designed for evidencing the bifurcation. 

Some parameters are easily optimized using the findings of
Sec.~\ref{hopping} and \ref{finite}. As discussed in
Sec.~\ref{sec:numerical-estimates}, cavity should be as short as
possible and in this respect a monolithic configuration would be
optimal. Then, crystal birefringence should be chosen as large as
possible [see Eq.~\eqref{eq:monolithic}]. However this may not be an
option as brifrigence is primarily used to achieve phase-matching in
the desired operating range. Of course, pump power should be as large
as possible, but our analysis has shown that there were modest returns
in increasing it much beyond that offered by currently available pump
lasers. The two parameters left for setup optimization are then the
signal and pump cavity finesses. Criterion~\eqref{eq:amonlefranc}
indicates that those finesses, and especially the pump finesse, should
be taken as large as possible.

However, cavities of very high finesse are critical to align and are
very sensitive to fluctuations. For example, a change of $\delta L$ of
the cavity length will induce a variation of $\delta \Delta = 4
\mathcal{F} \delta L/\lambda$ of the detuning, where $\lambda$ is the
optical wavelength. For $\lambda=1064$ nm and $\mathcal{F}=1000$, a
fluctuation $\delta L=5$ \AA \ of the cavity length will induce a
variation of the detuning $\delta\Delta=2$, corresponding to
full-width of the resonance. The length of the cavity must thus be
adjusted carefully and maintained constant to within less than a few
angstr\"{o}ms.

This stability problem is all the more critical when thermal effects
are taken into account. In our configuration, it has been shown that
thermal effects can induced complex dynamical
behaviors~\cite{suret00,suret01a} during which cavity length can be
spontaneously swept by a few nanometers. Such instabilities were
observed even at incident pump powers around 500 mW~\cite{suret00},
which for a pump finesse of 45 corresponds to an intracavity pump
power of about 6.5 W. For 4 W of incident pump power and a finesse of
500, intracavity pump power would reach $\sim$640 W at resonance and
thermal effects would then be a hundred times as strong. For
illustration purposes, let us recall that the only chaotic regime
reported so far in an TROPO was observed in a situation where cavity
length could not be made stationary~\cite{amon04}, although the
configuration was standard.

Another adverse influence of pump absorption in the crystal is that it
puts a limit on the highest finesse achievable. In our setup, for
example, pump absorbtion in the 1.5cm-long crystal is 2\% cm$^{-1}$
which implies that pump finesse cannot be higher than about 100 even
if perfectly reflecting mirrors were used. Except for crystals with
exceptional low absorption, high values of the pump cavity finesses
such as 500 appear to be completely unrealistic.

Apart from the previous remarks which hold for all systems featuring
an absorbing material enclosed in a resonant cavity, specific
restrictions on the finesses are to be considered for a TROPO, inside
which several fields must be simultaneously resonant. Generally cavity
finesses in an experimental TROPO setup are chosen so that
$\mathcal{F}_p \ll \mathcal{F}_s$. It ensures that numerous resonances
of the signal field are found inside the wider resonance of the pump
field, in other terms that several mode hops occur before OPO falls
below threshold. This allows one to only consider coincidences of
signal and idler cavity modes. For very high cavity finesses of the
pump, the mode with smallest signal detuning might very well be below
threshold because of a large pump detuning. In fact, the analysis of
part~\ref{hopping} should then be reworked and reformulated in terms
of coincidences of three resonances instead of two. It is not certain
that this more complicated problem admits solutions.

To conclude, an OPO designed in order to observe the Hopf bifurcation
should have a short cavity (e.g., monolithic setup), high
birefringence and pumping rate. The pump cavity finesse should be as
high as possible. Under these conditions, however, an OPO would be
very difficult to operate and to stabilize on an operating point. Even
then, the numerical analysis of Sec.~\ref{finite} indicates that
unstable zones would remain small compared to the range of operating
parameters. This casts serious doubts on the experimental
observability of the Hopf bifurcation of the monomode mean-field
model.

\section{conclusion}

In this paper we have raised the issue that contrary to what is
commonly assumed in theoretical studies of the dynamics of optical
parametric oscillators, signal detuning is not a fixed parameter but a
dynamical variable whose value results from a complex selection
process. Thus, detuning limitation due to mode hopping may affect the
occurrence of dynamical instabilities occurring at high values of the
detuning. As a simple example, we analyzed under which conditions the
Hopf bifurcation leading to periodic behavior in the monomode
mean-field model of a triply resonant OPO can be observed.

We first showed that signal detuning must reach a minimal value for
the bifurcation to occur, even at infinite pump power. We then
described how the mechanism of mode hopping restraints the values
signal detuning can take and we gave an analytical expression of the
maximal detuning allowed by mode hopping. By comparing the two bounds
so obtained, we showed that they are incompatible in many
configurations, showing that finite pump power is not necessarily the
limiting factor for observing this instability. This was confirmed in
numerical computations of instability domains carried out at a finite
pump power corresponding to our experimental configuration, in which
we observed that most of the unstable domains are inaccessible because
of mode hopping. We found that pump finesse is a critical parameter
that should be made as large as possible in order to observe the Hopf
bifurcation. However experimental setups designed for this purpose
should most certainly be extremely difficult to operate.

For the sake of simplicity, our derivation of the maximal detuning
allowed by mode hopping relied on two main hypotheses. One was that
signal and idler have frequencies close to 1:1 ratio but experience
different indices, as in our type-II experiments. Generalization to
other frequency ratios in the type-II configuration should be
straightforward and is not expected to modify the conclusions of the
present work. Type-I OPOs near frequency degeneracy have more complex
tuning properties and should be analyzed separately. The other
hypothesis was that parametric gain can be considered constant along a
frequency domain containing many longitudinal modes so that the lowest
possible detuning is always selected. If that is not the case, then
values of signal detuning higher than predicted may be obtained and
our analysis would have to be adapted. However, it does not seem
likely that our conclusions would be modified a lot.

An open question is whether conclusions drawn for the Hopf bifurcation
of the monomode mean-field model also hold for other theoretical
predictions of temporal or pattern-forming instabilities in doubly or
triply optical parametric oscillators. Indeed, such instabilities have
usually been predicted for relatively high values of the signal
detuning and/or of the pumping rate (see, e.g.,
Refs.~\cite{staliunas95,lugiato88,marte98,schwob98,tlidi00,oppo00}).
Thus, their observability may very well be also affected by detuning
limitation due to mode hopping rather than by finiteness of the pump
power.  A detailed comparison of instability thresholds in detuning
space with the maximum values allowed by mode hopping is in order for
these theoretical predictions so that their relevance for experimental
systems can be assessed. To conclude, mode hops are not only a
nuisance for stabilizing and tuning OPOs but also for using them as
tools to study complex dynamical behavior.

\section*{ACKNOWLEDGMENTS}
We are most grateful to S. Bielawski, D. Derozier, P. Suret, and
J. Zemmouri for stimulating discussions. The Centre d'\'Etudes et de
Recherches Lasers et Applications is supported by the Minist\`ere
charg\'e de la Recherche, the R\'egion Nord-Pas de Calais and the
Fonds Europ\'een de D\'eveloppement \'Economique des R\'egions.

\end{document}